\documentstyle{article}
\begin{document}
\title{
Quantum Entanglement  and  Quantum Computational Algorithms}
\author{Arvind \thanks{arvind@physics.iisc.ernet.in}}
\maketitle
\centerline{Department of Physics,} 
\centerline{Guru Nanak Dev University,}
\centerline{Amritsar, India 143 005}
\begin{abstract}
The existence of entangled quantum states gives extra power to
quantum computers over their classical counterparts. 
Quantum entanglement shows up qualitatively at the level of two
qubits.  We show that if no entanglement is envolved then 
whatever one can do with qubits can also be done with classical
optical systems.  We demonstrate that the one- and the two-bit
Deutsch-Jozsa algorithm does not require entanglement and
can be mapped onto a classical optical scheme. It is only for three 
and more input bits that the DJ algorithm  requires the implementation
of entangling transformations and in these cases it is impossible to
implement this algorithm classically.
\end{abstract}
\section{Introduction}
The realisation that quantum-mechanical systems have a very large 
in-built information processing ability and can hence be used to 
implement computational algorithms has aroused much interest 
recently~\cite{divin-sc-95,benn-pt-95,ch-sc-95,preskill-qph,benn-nat}.
The basic unit of quantum information is the \underline{qu}antum
\underline{bit} (qubit), which can be visualised as a quantum 
two-level system.  The implementation of quantum logic gates 
and circuits is based on the tenets of reversible logic and the fact 
that the two states of a qubit can be mapped onto logical 0 
and 1~\cite{bar-pr-95,divin-pr-95,monroe-prl-95,divin-roy-98}. 
Quantum mechanical realisation of logical operations can be used to 
achieve a computing power far beyond that of any classical 
computer~\cite{benn-ibm-73,feyn-the-82,benioff,deu-roy-85,deu-roy-89}. 
A few quantum algorithms have been designed and experimentally
implemented that perform certain computational tasks exponentially
faster than their classical counterparts. Three such algorithms
are, the Deutsch-Jozsa (DJ) algorithm ~\cite{deu-roy-92}, 
Shor's quantum factoring 
algorithm~\cite{shor-siam-97,ekert-revmod-96}, 
and Grover's rapid search algorithm~\cite{grover-prl-97}.

The existence of entangled states within quantum mechanics is 
one of the most striking features of the theory~\cite{schroedinger}. 
These states have the potential to show nontrivial
non-classical effects~\cite{ekert-amj-94,peres-prl,horo}.
It is now well recognised that the EPR paradox is based
on the existence of entangled states as only such states
exhibit correlations that are capable of violating
Bell's inequalities. It is rather interesting to note 
that entangled states play a crucial role in quantum computation 
as well, and it is the entanglement between
qubits that gives a quantum computer its inherent advantage.
A few researchers have discussed quantum algorithms which 
do not rely on quantum entanglement and yet are 
still intrinsically quantum in nature; however such 
computations are not based on qubits~\cite{lloyd-pra}.

We discuss in this  paper how entanglement prevents the 
mapping or realisation of a quantum computation using 
classical waves alone.  Consider the polarisation states 
of a classical light beam. These states are in one-to-one 
correspondence with the states of a qubit. All possible states 
can be realised by using one half-wave and two quarter-wave 
plates. One can thus pass from any desired polarisation
 state to all others in this way, i.e. all $U(2)\/$ transformations 
can be implemented using these gadgets~\cite{simon-pla-90}.
Therefore a single qubit has a classical analogue.  
On the other hand, it is not possible to map all the states of a 
two-qubit system onto the polarisation states of two light beams. 
The entangled states of the two qubits have no
classical counterpart. Therefore at the level of two qubits itself
the possibility of mapping a quantum computer onto 
classical optical fields breaks down.

To mathematically define a quantum-entangled state, consider a 
system composed of two parts and described by a density matrix 
$\rho\/$.  If the state of this  system can be written in terms 
of the states of its constituent systems in one of the following 
ways then it is \underline{not} entangled 
%\begin{mathletters}
\begin{eqnarray}
\rho&=& \rho^1 \otimes \rho^2 
\label{ent1}\\
\rho&=&\sum_{i}
\mbox{\boldmath $ p_i$\unboldmath}
\,\, \rho_i^1\otimes \rho_i^2
\quad \mbox{with}\quad \mbox{\boldmath $p_i$ \unboldmath}
 > 0.
\label{ent2}
\end{eqnarray}
%\end{mathletters}
In the case~(\ref{ent1}), the density matrix $\rho$ comprises 
of the tensor product density matrices $\rho_{1}\/$ and 
$\rho_{2}\/$, describing the constituent subsystems. Each 
system has a complete quantum description by itself and $\rho\/$ is
said to be strongly separable. On the other hand, in the 
case~(\ref{ent2}) the density matrix $\rho\/$ though not 
expressible as a tensor product of subsystem density matrices, 
is a positive sum of such tensor products. The positive coefficients
\boldmath$p_{i}\/$'s \unboldmath can be interpreted as probabilities
and therefore $\rho\/$ can be thought of as a classical mixture of 
strongly separable pieces. Such a density matrix is termed
weakly separable.  However, if the state $\rho\/$ is such that 
it cannot be expressed in either of the forms described in 
Eqns.~(\ref{ent1}) and~(\ref{ent2}), it is said to be entangled.
For separable states the correlations present can be 
given a classical meaning by interpreting the positive coefficients
\boldmath $p_i$'s \unboldmath as probabilities, an interpretation
which is not possible for entangled states. 

The central theme of this paper is to explore the scope of 
realising quantum computations on classical optical systems.
In this  context  we demonstrate the fact that the DJ algorithm 
for up to two input bits is classical, since  an explicit 
realisation of its implementation on a classical optical system 
based on polarisation is possible.  Further we show that
for more than two qubits the classical optical model fails 
since the algorithm essentially relies on quantum entanglement in
this case.  We stress that it is only at the level of three or 
more qubits that the DJ algorithm is  a ``truly quantum'' 
algorithm.
\section{The DJ algorithm}
The DJ algorithm was one of the first
algorithms that demonstrated the
power of quantum computers over classical ones~\cite{deu-roy-92}.
It determines whether a Boolean function $f\/$ is
constant or balanced. Classically, the algorithm requires
many function calls to solve the problem without error.
The quantum computer solves the problem using only a
single function call.

Consider an $n\/$-bit binary string $x\/$; a  
function $f\/$ can be defined on this $n\/$-bit
domain space to a 1-bit range space, with the
restriction that either the output is the same
for all inputs (the function is constant) or the
output is $0\/$ for half the inputs and $1\/$ for
the other half (the function is balanced).  All
the $2^n\/$ possible input strings are valid
inputs for the function ($f(x)=\{0,1\}\/$). 
In quantum computation, these
$n\/$-bit logical strings are in one-to-one
correspondence with the eigenstates of
$n\/$-qubits, and one can hence label the logical
string $x\/$ by the eigenstate $\vert x
\rangle\/$.  Classically, for an
$n\/$-bit domain space, one needs to compute the
function at least $2^{n/2}+1\/$ times  in order to
determine whether it is constant or balanced.
The DJ algorithm achieves this on a quantum computer using
only a single function
call~\cite{deu-roy-92,cleve-roy-98}.
The original algorithm required the implementation of
the function $f(x)\/$ encoded through a
unitary transformation on an extra qubit, 
along with
with the Hadamard transformation~\cite{deu-roy-92,cleve-roy-98}.

The Hadamard transformation for $n$-qubits 
is given by  
\begin{equation}
H^n \vert x \rangle = \sum_{y=0}^{2^n -1}
(-1)^{\oplus \sum_{j} x_j y_j} \vert y \rangle
\end{equation}
where $x_j\/$ and $y_j\/$ are the $j$th
entries of the $n$-bit strings $x$ and $y$ and $\oplus\/$ 
symbolises addition modulo 2.
The Hadamard transformation plays an important role
in many quantum computational algorithms and when
applied to the state $\vert 0 \rangle_{\rm n-bit}\/$
generates a superposition of all possible eigenstates 
of the $n$ qubit system. For a single qubit, the
Hadamard transformation reduces to
\begin{equation}
\begin{array}{c}
\vert 0\rangle \stackrel{H}{\rightarrow} 
\frac{\scriptstyle 1}{\scriptstyle \sqrt{2}}
(\vert 0 \rangle + \vert 1 \rangle) \\
\vert 1\rangle \stackrel{H}{\rightarrow} 
\frac{\scriptstyle 1}{\scriptstyle \sqrt{2}}
(\vert 0 \rangle - \vert 1 \rangle)
\end{array}
;\,
H = H^{-1}=\frac{1}{\sqrt{2}}
\left(\begin{array}{lr} 
{ 1} & {1}\\
{ 1} & {-1}
\end{array}
\right) 
\label{hadamard}
\end{equation}
The $n$-bit Hadamard transformation $H^n\/$ is non-entangling
in nature and is just a tensor product of n one-bit
transformations.

A modified scheme can be designed to solve
the $n$-bit Deutsch problem, using 
$n$ qubits alone~\cite{collins-pra-98}. 
Here, for every function $f\/$ a unitary
transformation is constructed,  such that its
action on the eigenstates of $n$-qubits is 
\begin{equation}
\vert x \rangle_{\mbox{n-bit}}
\stackrel{U_{f}}{\longrightarrow}
(-1)^{f(x)} \vert x \rangle_{\mbox{n-bit}}
\label{call-mech}
\end{equation}

Consider $n$ qubits, all in the state 
$\vert 0 \rangle\,$; a Hadamard transformation
$H^n\/$ converts this state to a linear
superposition of all $2^n\/$ eigenstates
with equal amplitudes and no phase differences. 
The unitary transformation $U_f\/$ (defined in
Eqn.~\ref{call-mech}) acting on this
state, introduces an $f$-dependent phase
factor in each eigenstate in the superposition.
 At this juncture, all
information about $f\/$ is encoded in the
quantum state of the $n$ qubits. A
Hadamard transformation $H^n\/$ is once
again applied in order to extract the function's
constant or balanced nature:  
\begin{eqnarray}
\vert 0 \rangle \stackrel{H^n}{\longrightarrow}
\sum_{x=0}^{2^n-1}\vert x \rangle
\stackrel{U_f}{\longrightarrow}
\sum_{x=0}^{2^n-1}
(-1)^{f(x)} \vert x \rangle 
\stackrel{H^n}{\longrightarrow} \quad\quad&&
\nonumber \\
\sum_{x=0}^{2^n-1}
\sum_{y=0}^{2^n-1}
(-1)^{f(x)} (-1)^{\oplus \sum_j x_j y_j} 
\vert y \rangle &&
\label{working-eq}
\end{eqnarray}
The final expression 
for the output state in Eqn.~\ref{working-eq}
has an amplitude $1\/$ for the state
$\vert 0 \rangle_{\mbox{n-bit}}\/$ for
a constant function and an amplitude $0\/$
for a balanced function.
The categorisation of the function  as constant
or balanced  through a single function call using
$n\/$ qubits, 
is shown pictorially
in Figure~1.
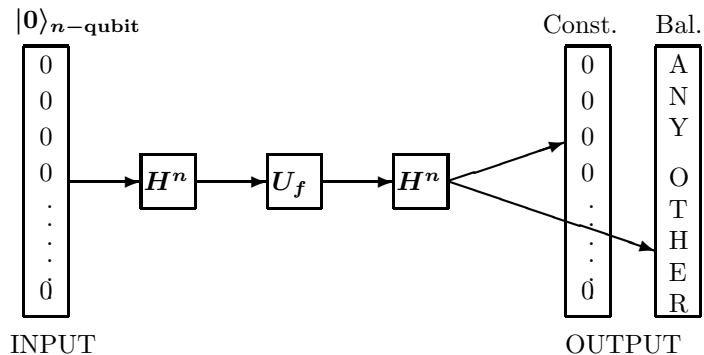
\begin{figure}[h]
\begin{center}
\unitlength=1.2mm
\begin{picture}(75,40)
\thicklines
\multiput(5,20)(14,0){3}{\vector(1,0){8}}
\put(47,20){\vector(3,1){13}}
\put(47,20){\vector(3,-1){23}}
\multiput(13,17)(14,0){3}{\line(1,0){6}}
\multiput(13,23)(14,0){3}{\line(1,0){6}}
\multiput(13,17)(14,0){3}{\line(0,1){6}}
\multiput(19,17)(14,0){3}{\line(0,1){6}}
\multiput(0,5)(60,0){2}{\line(1,0){5}}
\multiput(0,35)(60,0){2}{\line(1,0){5}}
\multiput(0,5)(60,0){2}{\line(0,1){30}}
\multiput(5,5)(60,0){2}{\line(0,1){30}}
\multiput(70,5)(0,30){2}{\line(1,0){5}}
\multiput(70,5)(5,0){2}{\line(0,1){30}}
\put(1.75,7){0}
\multiput(2.5,7)(0,2){6}{.}
\multiput(1.75,20)(0,4){4}{0}
\put(61.75,7){0}
\multiput(62.5,7)(0,2){6}{.}
\multiput(61.75,20)(0,4){4}{0}
\boldmath
\multiput(13.3,19)(28,0){2}{$H^n$}
\put(27.5,19){$U_f$}
\put(-1,37){$\vert 0 \rangle_{n-{\rm qubit}}$}
\unboldmath
\put(71.5,32){A}
\put(71.5,28.5){N}
\put(71.5,25){Y}
\put(71.5,19.5){O}
\put(71.5,16){T}
\put(71.5,12.5){H}
\put(71.5,9){E}
\put(71.5,5.5){R}
\put(-1.5,1){INPUT}
\put(60,1){OUTPUT}
\put(57.5,36.5){Const.}
\put(70,36.5){Bal.}
\end{picture}
\end{center}
\vspace{12pt}
\caption{The block diagram for the modified DJ algorithm}
\label{working}
\end{figure}
The number of functions for the $n$-bit Deutsch
problem is ${}^{N}C{}_{N/2} + 2\/$ (where
$N = 2^{n}\/$).   The experimental 
implementation of the modified DJ algorithm for 
$n\/$ bits requires the realisation of the  
unitary transformation corresponding to each
of these functions, and the $n$-bit Hadamard
transformation, on a physical system. 
There have been many  experimental 
implementations of the
DJ algorithm and its modified version, 
using NMR~\cite{ch-nat,j-jcp,lin,kav1,kav-dj,korean1,korean2}.
%%%%%%%%%%%%%%%
\section{Classical optical implementation of the DJ algorithm}
We now proceed towards analysing this algorithm
for 1 and 2 bits in detail, with a view to
exploring the possibility of realising it on a 
classical optical system. Let us consider the following 
classical system: a  monochromatic light beam propagating in 
a given direction with a pure polarisation. The
polarisation states of such a beam are in one-to-one 
correspondence with the states of a two-level system 
and this beam can therefore be visualised as a qubit. 
It is also well known that all unitary 
transformations on the polarisation
states can be performed.
Consider a birefringent plate with its thickness adjusted
to introduce a phase difference of $\eta\/$ between the
$x\/$ and $y\/$ components of the electric field and 
whose slow axis
makes an angle $\phi\/$ with the $x\/$ axis.
The unitary operator corresponding to this plate is given by
\begin{equation}
U(\eta,\phi) =
\left(\begin{array}{cc} \cos\phi & -\sin \phi\\
                         \sin\phi & \cos \phi\end{array} \right)
\left(\begin{array}{cc} e^{i\eta/2} & 0\\ 0 & e^{-i \eta/2}
\end{array} \right)
\left(\begin{array}{cc} \cos\phi & \sin \phi\\
                         -\sin\phi & \cos \phi\end{array} \right)
\end{equation}
For the choice $\eta=\pi\/$ it becomes a half-wave plate and 
we denote it by $H_{\phi}\/$, while for $\eta=\pi/2\/$ it becomes
a quarter-wave plate and we denote it by $Q_{\phi}\/$.
It has been  shown that all $SU(2)\/$ transformations can be 
realised on the polarisation states by taking two quarter-wave
plates and one half-wave plate with suitable choices of angles of
their slow axes with the $x\/$ axis.  We will henceforth refer 
to this device capable of implementing $SU(2)\/$ transformations
as Q-H-Q and a detailed discussion is found in~\cite{simon-pla-90}.

Further, let us map the $x\/$ polarisation state to logical $1\/$ 
and the $y\/$ polarisation state to logical $0\/$.  With this 
mapping we can proceed to work with this system as a qubit.
Since this system comprises essentially of
classical elements, we call it a ``classical qubit''.

\noindent{\bf One-bit case:}
For the purpose of implementing the one-bit DJ algorithm 
on a ``classical qubit'', we need to  realise the
Hadamard transformation $H\/$ which corresponds to
an anti-clockwise rotation of polarisation by $45^0\/$ 
in the $x-y\/$ plane and the four $U_f\/$ transformations 
corresponding to the four possible functions.  These are certain
$SU(2)\/$ transformations and therefore are
implementable using the Q-H-Q device discussed above.
The sequence of optical operations 
is detailed in Figure~2.
As an example, consider the third 
unitary transformation 
\begin{equation}
U_{3}^{\rm 1-bit} = \left(
\begin{array}{cc}
1 & 0 \\
0 & -1
\end{array}
\right)
\end{equation}
which when realised on a ``classical qubit'' would 
leave the  $y$ polarisation unaffected while the $x$ 
polarisation picks up a phase factor of $e^{i \pi}\/$. This 
transformation is achievable by a single half-wave plate.

\def\arvbox{\multiput(0,0)(8.5,0){2}{\line(0,1){14}}
\multiput(0,0)(0,14){2}{\line(1,0){8.5}} }
\begin{figure}[h]
\unitlength=0.9mm
\begin{picture}(100,30)(-15,0)
\boldmath
\thicklines
\multiput(0,0)(0,16){1}{
\multiput(20,0)(25,0){4}{\arvbox}
\multiput(3.5,7)(25,0){4}{\line(1,0){16.5}}
\put(5,8.5){y-pol.}
\multiput(0,0)(50,0){2}
{
\put(22,7.5){45$^0$}
\put(20.4,4){ROT}
}
\put(98,8.5){$\hat{y}$}
\put(95.5,4.2){POL}
}
\put(47,6){$U_{f}$}
\put(104,10){
\begin{tabular}{l}
\\
\\
Intensity\\
Measurement\\
\end{tabular}}
\end{picture}
\label{one-bit}
\vspace*{24pt}
\caption{Optical realisation of 1-bit DJ algorithm.
If the resultant polarisation after the second $45^0\/$ rotation
is along $\hat{y}\/$ then the function is constant and if it is along
$x\/$ the function is balanced. Since these two polarisation states
are orthogonal it is possible to distinguish them with certainty
by placing a $y$ polarizer after the second $45^0\/$ 
rotation, followed by an intensity measurement.} 
\end{figure}
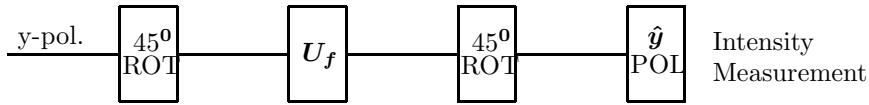
The outcome of this exercise is to show 
that there is nothing quantum
about this procedure, and that the 
concept of a qubit can be 
mapped onto the polarisation states of a light beam.

\noindent{\bf Two-bit case:}
For implementing the two-bit DJ algorithm, we
consider two beams of monochromatic light
representing two ``classical qubits''. The 
Hadamard transformation $H^2\/$
is again a rotation of the polarisation of both the beams 
in the anti-clockwise direction
by an angle of $45^0\/$. All the eight $U_f^{\rm 2-bit}\/$ 
transformations corresponding to 
the eight functions turn out to be factorisable as a direct 
product of two $SU(2)\/$ transformations,
one acting on each qubit.
\begin{equation}
U_f^{\rm 2-bit}=U_f^1\otimes U_f^2\quad
 {\rm with}\quad U_f^1,U_f^2 \in SU(2)
\end{equation}
More specifically we give in Eqn.~\ref{two-bit-uf} 
the decomposition of four of 
the eight functions. The other four transformations differ 
from these four by an overall phase factor of $\pi\/$ and
therefore can be similarly factorised.
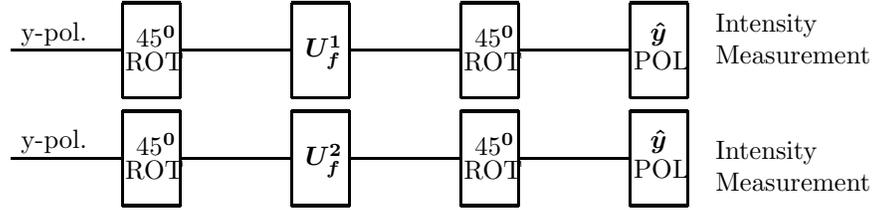
\begin{figure}[h]
\unitlength=0.9mm
\begin{picture}(120,60)(-15,0)
\boldmath
\thicklines
\multiput(0,0)(0,16){2}{
\multiput(20,0)(25,0){4}{\arvbox}
\multiput(3.5,7)(25,0){4}{\line(1,0){16.5}}
\put(5,8.5){y-pol.}
\multiput(0,0)(50,0){2}
{
\put(22,7.5){45$^0$}
\put(20.4,4){ROT}
}
\put(98,8.5){$\hat{y}$}
\put(95.5,4.2){POL}
}
\put(47,6){$U^2_{f}$}
\put(47,22){$U^1_{f}$}
\put(104,14){
\begin{tabular}{l}
Intensity\\
Measurement\\
\\
\\
Intensity\\
Measurement\\
\end{tabular}}
\end{picture}
\vspace*{24pt}
\label{two-bit-optical}
\caption{Optical realisation of the DJ algorithm
for two qubits. Similar to the one-bit case, the
polarisers along $\hat{y}\/$
are placed after the final Hadamard
transformation to project out the $\vert 0 0 \rangle\/$
state. The presence of light signal in both the
beams indicates a constant function while the absence
of signal in either of the beams indicates a balanced
function.}
\end{figure}

\begin{equation}
\begin{array}{lll}
\!\!\!
\!\!\!
\!\!\!
\!\!\!\!\!\
U_{1}^{\rm 2-bit} = \left( \begin{array}{cccc}
1 & 0 & 0 & 0 \\
0 & 1 & 0 & 0 \\
0 & 0 & 1 & 0 \\
0 & 0 & 0 & 1
\end{array}
\right)
&= \left( \begin{array}{rr}
1 & 0 \\
0 & 1
\end{array}
\right)
\otimes
\left( \begin{array}{rr}
1 & 0 \\
0 & 1
\end{array}
\right)& \mbox{\rm No Operation}\nonumber \\
\!\!\!
\!\!\!
\!\!\!\!\!\!\!\!\
U_{2}^{\rm 2-bit} = \left( \begin{array}{cccc}
1 & 0 & 0 & 0 \\
0 & -1 & 0 & 0 \\
0 & 0 & 1 & 0 \\
0 & 0 & 0 & -1
\end{array}
\right)
&= \left( \begin{array}{rr}
1 & 0 \\
0 & 1
\end{array}
\right)
\otimes
\left( \begin{array}{rr}
1 & 0 \\
0 & -1
\end{array}
\right) &
\mbox{\rm HWP on 1st beam}
 \nonumber \\
\!\!\!
\!\!\!
\!\!\!\!\!\!\!\!\
U_{3}^{\rm 2-bit} = \left( \begin{array}{cccc}
1 & 0 & 0 & 0 \\
0 & 1 & 0 & 0 \\
0 & 0 & -1 & 0 \\
0 & 0 & 0 & -1
\end{array}
\right)
&= \left( \begin{array}{rr}
1 & 0 \\
0 & -1
\end{array}
\right)
\otimes
\left( \begin{array}{rr}
1 & 0 \\
0 & 1
\end{array}
\right) 
&\mbox{\rm HWP on 2nd beam}
\nonumber \\
\!\!\!
\!\!\!
\!\!\!\!\!\!\!\!\
U_{4}^{\rm 2-bit} = \left( \begin{array}{cccc}
1 & 0 & 0 & 0 \\
0 & -1 & 0 & 0 \\
0 & 0 & -1 & 0 \\
0 & 0 & 0 & 1
\end{array}
\right)
&= \left( \begin{array}{rr}
1 & 0 \\
0 & -1
\end{array}
\right)
\otimes
\left( \begin{array}{rr}
1 & 0 \\
0 & -1
\end{array}
\right) &\mbox{\rm HWP on both beams}
\end{array}
\label{two-bit-uf}
\end{equation}

To implement these transformations on two ``classical qubits'' 
requires insertion of a half wave plate in one or both beams and
can be achieved in a straightforward way.
Hence the implementation of $U_f^{\rm 2-bit}\/$ can be achieved
by  applying the required $SU(2)\/$ transformation on each
qubit separately. The schematic diagram of 
this implementation is shown in
Figure~3.
\section{Entanglement at the three-qubit level}  
We now turn to the case of three input qubits. 
In this case too the Hadamard 
transformation $H^{\rm 3}\/$ corresponds to a $45^0\/$ rotation of 
the polarisations of each of the three qubits and can be 
implemented easily.
The functions $U_f^{\rm 3-bit}\/$ are 72 in number and 
any physical implementation would require
a prescription to implement all of them on a system of  
three qubits. Consider for example, a particular balanced 
function given by the $8 \times 8\/$
diagonal matrix 
\begin{equation}
U_f=\left(\begin{array}{rrrrrrrr}
1&0&0&0&0&0&0&0\\
0&-1&0&0&0&0&0&0\\
0&0&1&0&0&0&0&0\\
0&0&0&1&0&0&0&0\\
0&0&0&0&-1&0&0&0\\
0&0&0&0&0&-1&0&0\\
0&0&0&0&0&0&-1&0\\
0&0&0&0&0&0&0&1
\end{array} \right)
\end{equation}
This simple looking $SU(8)\/$
diagonal matrix cannot be written as a direct
product of $SU(2)\/$ matrices acting on individual qubits!.
This transformation is entangling in nature and possesses the
capacity to generate entangled states from non-entangled ones.
To see this clearly, let us consider the action of this matrix
on a simple un-entangled initial state 
\begin{equation}
U_f \frac{1}{2\sqrt{2}}\left[
\left(\begin{array}{r}1\\1\end{array}\right)
\otimes\left(\begin{array}{r}1\\1\end{array}\right)
\otimes\left(\begin{array}{r}1\\1\end{array}\right)
\right] = \frac{1}{2\sqrt{2}}\left(\begin{array}{r}
1\\1\\-1\\1\\-1\\-1\\-1\\1
\end{array} 
\right)
\end{equation}
The entanglement of this final pure three-qubit state can be
demonstrated by computing the reduced density matrix for
the last two qubits, by taking a partial trace over the first qubit
\begin{eqnarray}
\rho^{2-3}&=&\frac{1}{4}
\left(
\begin{array}{cccc}
1&0&-1&0\\
0&1&0&1\\
-1&0&1&0\\
0&1&0&1\\
\end{array}
\right)
\nonumber \\
\mbox{\rm where} (\rho^{2-3})^2 &\neq& \rho^{2-3}
\end{eqnarray}
The result $(\rho^{2-3})^2 \neq \rho^{2-3}\/$ implies
that the reduced density matrix is mixed, proving the
entangled nature of the overall state of the 3-qubit system.
The question now arises as to how do we realise such an 
entangling transformation for our ``classical qubit'' system?
It turns out that there is no way we can realise such 
transformations on ``classical qubits''.
It is to be noted that 
there are $SU(2)\otimes SU(2) \otimes 
SU(2)\/$ worth of states for the  
system of three 
``classical qubits''  under consideration and this does not 
contain the entangled states. 
As shown above, entangling transformations on the other hand,
generate entangled states 
from non-entangled ones and therefore
there is no
possibility of being able to construct and implement 
them on this system of ``classical qubits''.
Hence in  order to implement the three bit DJ
algorithm we need a {\em real quantum-mechanical three-qubit} 
system.  Such implementations have been discussed in the
context of liquid state NMR quantum 
computing~\cite{ch-nat,j-jcp,lin,kav1,kav-dj,korean1,korean2}.
\section{Concluding Remarks}
A complete mapping of 
a qubit exists in the classical world of polarisation optics. 
We can consider $SU(2)\/$ worth of pure polarisation
states of a monochromatic beam  of light as a qubit.
All unitary operations required to implement logical operations
being $SU(2)\/$ transformations, can be implemented through Q-H-Q.
We have shown that the one-bit DJ algorithm can be readily
implemented on this system.
More generally, we can say that every operation which can be conceived
of for a single qubit, can be executed on this classical system.

We have shown that 
for more than one qubit the ``classical qubit'' system can work 
for those computations which do not involve entanglement and
for such cases there might be an advantage over the ordinary
binary computers. In particular for the two-bit DJ algorithm,
since it does not involve quantum entanglement in its 
implementation for two bits, we are able to realise 
it on our ``classical
qubits''. This algorithm for the two-bit case solves the
problem with only one function call as opposed to the 
ordinary classical algorithm requiring three  function calls.
Hence even when this algorithm  allows 
a realisation based on the ``classical
qubits'' it outperforms the algorithm based on ordinary binary 
logic. Therefore, using a pair of 
``classical qubits'' has some advantage!

Finally we illustrate that an
algorithm becomes a true quantum algorithm only when it involves
quantum entanglement at some stage of its implementation, 
otherwise it is implementable on a set of ``classical 
qubits''. The  DJ algorithm for three qubits has been shown to 
involve entangled states for its implementation. Therefore,
it becomes impossible to realise its implementation using ``classical
qubits''.

\vspace*{6pt}
\noindent{\bf Acknowledgements:} I thank my
collaborators Kavita Dorai, Anil Kumar,
N.~Mukunda and R.~Simon for useful discussions.

%\end{references}
\end{document}